\documentclass[sigconf,natbib=true,anonymous=false]{acmart}

\makeatother

\copyrightyear{2025}
\acmYear{2025}
\setcopyright{cc}
\setcctype{by-nc-sa}
\acmConference[ICTIR '25]{Proceedings of the 2025 International ACM SIGIR Conference on Innovative Concepts and Theories in Information Retrieval (ICTIR)}{July 18, 2025}{Padua, Italy}
\acmBooktitle{Proceedings of the 2025 International ACM SIGIR Conference on Innovative Concepts and Theories in Information Retrieval (ICTIR) (ICTIR '25), July 18, 2025, Padua, Italy}
\acmDOI{10.1145/3731120.3744613}
\acmISBN{979-8-4007-1861-8/2025/07}

\usepackage{enumitem}
\usepackage{xspace}
\usepackage{bbm}
\usepackage{algorithm}
\usepackage{algpseudocode}
\usepackage{balance}

\algrenewcommand\algorithmicindent{1.0em}
\algnewcommand\Input{\item[\textbf{Input:}]}
\algnewcommand\Output{\item[\textbf{Output:}]}

\usepackage{amsmath}
\DeclareMathOperator*{\argmax}{arg\,max}

\begin{document}
\title{Distillation and Refinement of Reasoning in Small Language Models for Document Re-ranking}

\author{Chris Samarinas}
\affiliation{%
  \institution{University of Massachusetts Amherst}
  \city{Amherst}
  \state{MA}
  \country{United States}
}
\email{csamarinas@cs.umass.edu}

\author{Hamed Zamani}
\affiliation{%
  \institution{University of Massachusetts Amherst}
  \city{Amherst}
  \state{MA}
  \country{United States}
}
\email{zamani@cs.umass.edu}

\newcommand{\ranker}{InteRank}

\begin{abstract}
We present a novel approach for training small language models for reasoning-intensive document ranking that combines knowledge distillation with reinforcement learning optimization. While existing methods often rely on expensive human annotations or large black-box language models, our methodology leverages web data and a teacher LLM to automatically generate high-quality training examples with relevance explanations. By framing document ranking as a reinforcement learning problem and incentivizing explicit reasoning capabilities, we train a compact 3B parameter language model that achieves state-of-the-art performance on the BRIGHT benchmark. Our model ranks third on the leaderboard while using substantially fewer parameters than other approaches, outperforming models that are over 20 times larger. Through extensive experiments, we demonstrate that generating explanations during inference, rather than directly predicting relevance scores, enables more effective reasoning with smaller language models. The self-supervised nature of our method offers a scalable and interpretable solution for modern information retrieval systems.

\end{abstract}



\maketitle

\begin{figure*}[t]
    \centering
    \includegraphics[width=0.85\textwidth]{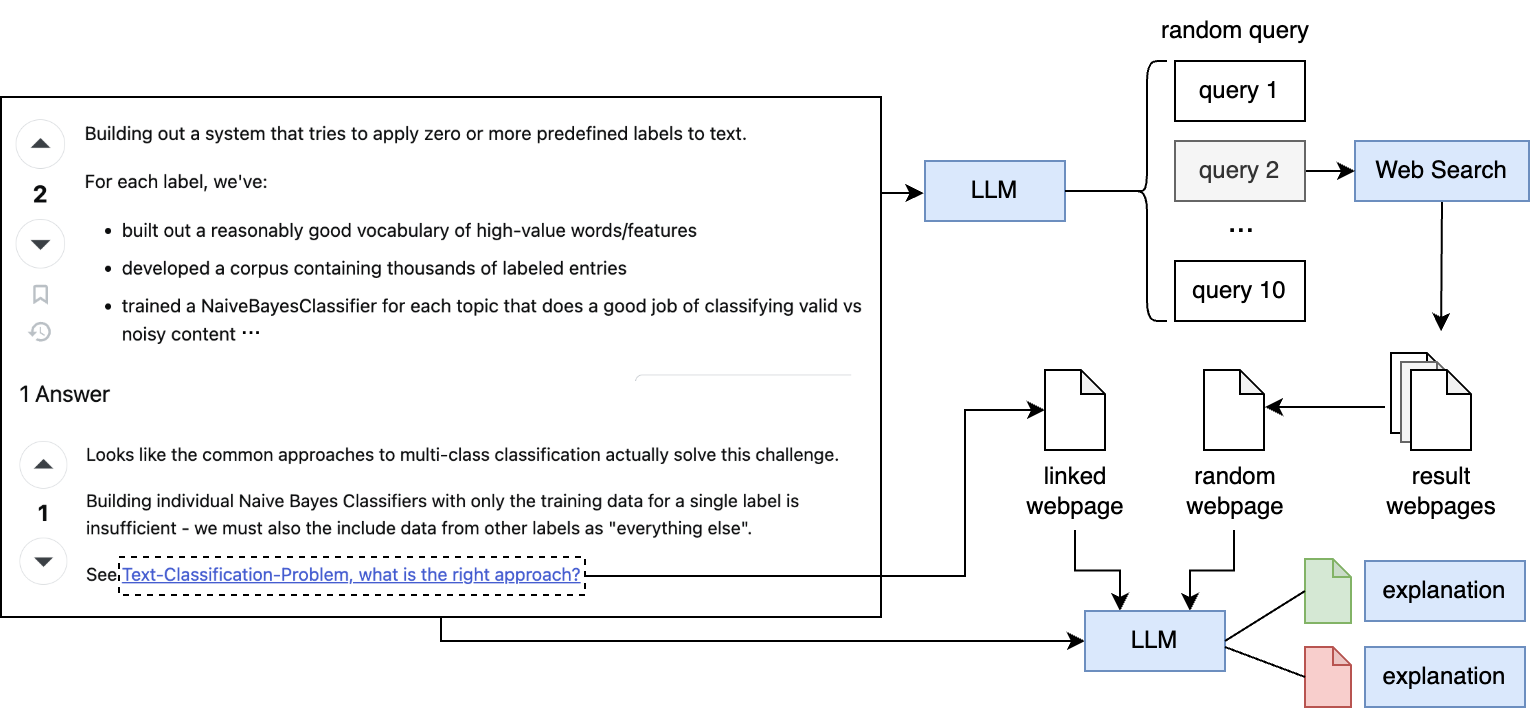}
    \caption{The automated synthetic data annotation process based on Stackexchange question-answer pairs.}
    \label{fig:data-generation}
\end{figure*}

\section{Introduction}
Search engines and retrieval-augmented generation systems increasingly face queries that require complex reasoning and multi-step synthesis and analysis. They demand a deeper understanding of the query and the documents to identify the connections between them. For example, finding documentation for a coding error requires understanding program logic and syntax, and identifying economic case studies that share underlying theoretical principles demands sophisticated domain knowledge and analytical reasoning \cite{su2024bright}. Traditional approaches to training ranking models for such complex tasks often rely on expensive human annotations to provide relevance judgments and explanations. In contrast, we present a framework that automatically generates its own training signal by leveraging existing question-answer pairs on the Web.

Although neural ranking models have made significant progress in recent years \cite{nogueira2019passage,nogueira2019multi,devlin2018bert,samarinas-etal-2021-improving,hypencoder} and led to substantial performance gains on standard benchmarks such as MS MARCO \cite{DBLP:journals/corr/NguyenRSGTMD16} and the TREC Deep Learning (DL) Track \cite{craswell2021trec,craswell2020overview}, we observe that they often struggle with reasoning-intensive queries that demand deeper understanding and explicit justification of relevance decisions. For instance, state-of-the-art dense retrievers that achieve strong performance on TREC DL show significant degradation on reasoning-intensive queries, with the best models achieving only about 18\% nDCG@10 on the BRIGHT benchmark \cite{su2024bright}--a recent benchmark designed for reasoning-intensive ranking tasks. We argue that ranking models must engage in deliberate \emph{reasoning} to bridge the gap between query intent and document content.

Recent work has suggested that large language models with tens of billions of parameters can effectively serve as zero-shot rerankers \cite{sun2023chatgpt,pradeep2023rankzephyr,pradeep2023rankvicuna}, demonstrating strong reasoning capabilities across diverse domains. However, deploying these models at scale remains challenging due to their computational requirements and latency constraints. While smaller models offer practical advantages, they typically lack the sophisticated reasoning abilities of their larger counterparts. Recent LLMs such as DeepSeek R1 \cite{deepseekr1} have demonstrated that encouraging models to learn explicit reasoning strategies and leveraging inference-time compute for step-by-step analysis can significantly improve performance on complex tasks. While this has been demonstrated for language modeling and generation tasks, exploring these principles in retrieval remains understudied. Our work shows that by decomposing document relevance assessment into explicit reasoning steps and optimizing for high-quality explanations, we can achieve strong performance even with relatively compact models. 

In more detail, our work introduces a framework for distilling and refining reasoning capabilities in small language models for reasoning-intensive ranking. Our approach does not require any manually labeled data for training; instead, we perform a diverse data scraping approach from the Web for collecting reasoning intensive questions and a pseudo-labeling approach using a \textit{teacher} LLM (with 70B parameters), resulting in a dataset with 20K examples. We then introduce a knowledge distillation approach that helps a compact \textit{student} LLM (with 3B parameters) to mimic the reasoning and labeling capability of the teacher. Subsequently, we introduce a reinforcement learning approach that refines these reasoning capabilities by rewarding high-quality explanations and accurate relevance predictions. 

Through this approach, we demonstrate that our efficient 3B parameter model achieves performance comparable to 70B+ parameter models on reasoning-intensive ranking tasks. Most notably, our model ranks third on the BRIGHT benchmark leaderboard and is the first effective ranking model under 8B parameters, with the only models achieving better performance being a 70B zero-shot ranker using GPT-4 for query reformulation and JudgeRank, an ensemble of three LLMs (8B, 70B, and 405B parameters). Our 3B parameter model outperforms all other baseline methods on the BRIGHT benchmark, including recent approaches like Reason-to-Rank \cite{r2r2024} which uses an 8B parameter model, while using almost three times fewer parameters and avoiding complex query rewriting or multi-step prompting strategies. We release our code and data for improved reproducibility.\footnote{\url{https://github.com/algoprog/InteRank}}

\section{\ranker{}}


In this section, we present the training methodology for \ranker{}, a compact LLM for reasoning-intensive ranking. We leverage the reasoning capabilities of a large \textit{teacher} LLM to train a compact \textit{student} LLM that can both effectively re-rank documents and explain its decisions. The key insight is that by decomposing the ranking process into explicit reasoning steps and dedicating inference-time compute to step-by-step analysis, we can achieve superior performance compared to approaches that attempt to directly predict relevance scores. By training on synthetic explanations from a \textit{teacher} LLM and optimizing for high-quality reasoning paths with reinforcement learning, we can effectively transfer reasoning capabilities to compact LLMs without requiring human-annotated data.



\subsection{Model Architecture}

We adopt a two-stage ranking architecture that is common in modern search systems: an efficient first-stage retrieval followed by more expressive reranking model capable of reasoning.

\textbf{First-stage Retrieval:} A lightweight sparse or dense retrieval model is used to retrieve potentially relevant documents from the corpus. To better understand the impact of retrieval quality on the final ranking performance, we experiment with various retrievers, including BM25 \cite{robertson1995okapi} and dense embedding models (see section~\ref{experiments}). We retrieve the top 100 documents for re-ranking. While optimizing the first-stage retriever is important, in this work we focus on improving the second-stage re-ranking component.

\textbf{Second-stage Re-ranking:} Various learning-to-rank models, from traditional feature-based \cite{ltr,burges2010ranknet} to transformer-based cross-encoder models \cite{nogueira2019passage,nogueira2019multi,gao2021rethink}, have been used for reranking. We aim at training a reranking model for effective reasoning-intensive tasks. To do so, we train a language model that takes a query-document pair at a time and generates some reasoning to analyze and describe whether and how the provided document is relevant. These reasoning steps are then followed by a discrete relevance label as the final generation token. This relevance label is either 0 (i.e., non-relevant), 1 (partially relevant), or 2 (highly relevant). This stage is crucial for complex reasoning tasks, as it allows deeper analysis of document content in relation to the query intent. Since the scores produced by our reranker are discrete, many documents are assigned the same relevance score, and we cannot distinguish them in ranking. Therefore, we employ a hybrid scoring strategy that combines the generated discrete reranking score with the (dot product) retrieval score produced in the first-stage retrieval. In fact,
\begin{equation*}
\text{score}(q, d) = \text{retrieval score}(q, d) + \alpha \cdot\text{ reranking score}(q, d)
\end{equation*} 
where $q$ and $d$ denote query and document and $\alpha \in \mathbb{R}^+$ is a hyper-parameter controlling the impact of re-ranking score. $\alpha$ is expected to be a relatively high number $\gg 1$. ($\alpha=100$ in our experiments).


\subsection{Model Optimization}

Our training process combines knowledge distillation \cite{hinton2015distillingknowledgeneuralnetwork} with reinforcement learning (RL). Following recent work showing the benefits of incentivizing explicit reasoning capabilities through RL \cite{deepseekr1}, we structure our approach to encourage the development of effective reasoning patterns while maintaining computational efficiency. The process consists of three phases:

\textbf{1. Synthetic Data Generation:} High-quality training data is crucial for developing models that can handle diverse reasoning patterns. However, obtaining human annotations for reasoning-intensive ranking is expensive and time-consuming. We address this challenge through an automated data generation process that leverages existing question-answer pairs from social websites like StackExchange. Our data generation pipeline, summarized in Algorithm~\ref{alg:data-generation} and Figure~\ref{fig:data-generation}, starts with a seed set of query-answer pairs $\mathcal{C}$. In our experiments, we sampled 20K pairs round robin from 186 different communities on StackExchange. To prevent data contamination, we ensure that any examples appearing in the BRIGHT benchmark from StackExchange are excluded from our training data. For each answer, we extract linked documents (hyperlinks) that potentially contain supporting evidence, establishing an initial set of query-document pairs. To increase diversity and to source potential negative documents, we use a \textit{teacher} LLM to generate related queries and retrieve additional documents through web search using the Brave Search API. The teacher model is then instructed to generate an explanation and a discrete relevance label for each query-document pair, creating a distillation dataset. This approach naturally captures diverse reasoning patterns since the teacher model must explain how different types of evidence support or fail to support answers across technical domains - from code analysis to scientific explanations. The explanations demonstrate different forms of reasoning like logical deduction, causal analysis, and domain-specific technical reasoning. The next two phases are summarized in Algorithm~\ref{alg:training}.

\textbf{2. Knowledge Distillation:} We first transfer knowledge from a large zero-shot teacher model to a more compact student model through supervised fine-tuning. We use Llama 3.2 3B \cite{grattafiori2024llama3herdmodels} as our base student model and as mentioned earlier, Llama 3.3 70B is used as our teacher model. The objective is to maximize the log likelihood of teacher-generated outputs:
\begin{align}
\theta^1 &= \argmax_\theta \mathbb{E}_{(q,d,e,l)\sim D_{\text{synth}}} \left[\log p_\theta(e,l|q,d)\right]
\end{align}
where $e$ and $l$ denote an explanation and a discrete relevance label. $\theta^1$ is the trained student model parameters after knowledge distillation. This phase helps the student model learn some initial reasoning patterns.

\textbf{3. Reinforcement Learning:} While distillation helps transfer basic reasoning patterns from the teacher, it is limited to imitating a single explanation path per example. In practice, there may be multiple valid ways to reason about document relevance. The reinforcement learning (RL) phase enables exploration of diverse reasoning strategies through sampling, with the reward model providing feedback to identify the most effective explanations. For each query-document pair, we sample $k=8$ outputs from the model being trained (i.e., starting from the student model from Step 2) and evaluate them using a reward model. We observed that the reward values can have very high variance, and they heavily depend on query complexity and domain. For this reason, we use relative reward values after max-min normalization for each set of outputs $\hat{y}$ for a given query-document input $(q,d)$:
\begin{equation}
\overline{\mathcal{R}}(q,d,\hat{y}) = \frac{\mathcal{R}(q,d,\hat{y}) - \min(\mathcal{R})}{\max(\mathcal{R}) - \min(\mathcal{R})}
\end{equation}

\noindent
where $\min(\mathcal{R})$ and $\max(\mathcal{R})$ are the minimum and maximum reward values for the given query-document input pair. High-quality output samples $\hat{y}_j$ are then selected using a threshold $\tau$: 
\begin{equation}
D_t = \{(q,d,\hat{y}_j,\overline{\mathcal{R}}) : \overline{\mathcal{R}}(q,d,\hat{y}_j) \geq \tau\}
\end{equation}
\noindent
The model parameters are updated using scaled rewards to further emphasize higher-reward outputs:

\begin{equation}
\theta^{t+1} = \argmax_\theta \mathbb{E}_{(q,d,\hat{y},\overline{\mathcal{R}})\sim D_t} [(\overline{\mathcal{R}}(q,d,\hat{y}))^m \log p_\theta(\hat{y}|q,d)]
\vspace{-0.17cm}
\end{equation}

\begin{algorithm}[t!]
\small
\caption{Synthetic Data Generation for Ranking}
\label{alg:data-generation}
\begin{algorithmic}[1]
\Input Teacher LLM $\mathcal{T}$, query-answer pairs $\mathcal{C} = \{(q_i, a_i)\}_{i=1}^N$
\Output Synthetic dataset $D_{\text{synth}}$
\State Initialize $D_{\text{synth}} \gets \emptyset$
\For{each $(q,a) \in \mathcal{C}$}
    \State Extract linked documents $D_{\text{linked}}$ from $a$
    \For{each $d \in D_{\text{linked}}$}
        \State $(e,l) \gets \mathcal{T}(q,d)$ \Comment{Generate explanation and label}
        \State $D_{\text{synth}} \gets D_{\text{synth}} \cup \{(q,d,e,l)\}$
    \EndFor
    \State $Q_{\text{gen}} \gets \mathcal{T}(q,a,D_{\text{linked}})$ \Comment{Generate related queries}
    \State Sample random $q' \sim Q_{\text{gen}}$
    \State $D_{\text{web}} \gets \text{WebSearch}(q')$ \Comment{Get top-10 results}
    \State Sample random $d \sim D_{\text{web}}$
    \State $(e,l) \gets \mathcal{T}(q',d)$
    \State $D_{\text{synth}} \gets D_{\text{synth}} \cup \{(q',d,e,l)\}$
\EndFor
\State \Return $D_{\text{synth}}$
\end{algorithmic}
\end{algorithm}

\begin{algorithm}[t!]
\small
\caption{LLM alignment for ranking}
\label{alg:training}
\begin{algorithmic}[1]
\Input Student LLM $\mathcal{M}_\theta$, reward model $\mathcal{R}$, synthetic dataset $D_{\text{synth}}$
\Output Trained model parameters $\theta^{T+1}$
\State $\theta^1 \gets \argmax_\theta \mathbb{E}_{(q,d,e,l)\sim D_{\text{synth}}} [\log p_\theta(e,l|q,d)]$
\For{$t = 1$ to $T$}
    \For{each $(q,d,l)$ in training data}
        \State Sample $Y_{q,d} = \{\hat{y}_j\}_{j=1}^k \sim M_{\theta^t}(q,d)$ \Comment{$k=8$ samples}
        \State Compute rewards $\mathcal{R}(q,d,\hat{y}_j)$ for all $\hat{y}_j$
        \State Normalize rewards: $\overline{\mathcal{R}} = \frac{\mathcal{R} - \min(\mathcal{R})}{\max(\mathcal{R}) - \min(\mathcal{R})}$
    \EndFor
    \State $D_t = \{(q,d,\hat{y}_j,\overline{\mathcal{R}}(q,d,\hat{y}_j)) : \overline{\mathcal{R}}(q,d,\hat{y}_j) \geq \tau\}$
    \State $\theta^{t+1} \gets \argmax_\theta \mathbb{E}_{(q,d,\hat{y},\overline{\mathcal{R}})\sim D_t} [\overline{\mathcal{R}}^m \log p_\theta(\hat{y}|q,d)]$
    \Comment{$m=3$}
\EndFor
\State \Return $\theta^{T+1}$
\end{algorithmic}
\end{algorithm}

\begin{table*}[ht]
\centering
\footnotesize
\caption{Performance (nDCG@10) of different retriever and reranker combinations on the BRIGHT benchmark. Our 3B parameter model \ranker{}, matches or exceeds the performance of the 70B teacher LLM, with explanations being crucial for effectiveness (see "w/o expl." ablations). Adding domain-specific relevance definitions (marked with "+ instruct") further improves performance. The symbol * indicates statistical significance (paired t-test, p < 0.05) compared to all baseline models.}
\begin{tabular}{l|l|ccccccc|cc|ccc|c}
\hline
& & \multicolumn{7}{c|}{\textbf{StackExchange}} & \multicolumn{2}{c|}{\textbf{Coding}} & \multicolumn{3}{c|}{\textbf{Theorem-based}} & \\
\textbf{Retriever} & \textbf{Re-ranker} & \textbf{Bio.} & \textbf{Earth.} & \textbf{Econ.} & \textbf{Psy.} & \textbf{Rob.} & \textbf{Stack.} & \textbf{Sus.} & \textbf{Leet.} & \textbf{Pony} & \textbf{AoPS} & \textbf{TheoQ.} & \textbf{TheoT.} & \textbf{Avg.} \\
\hline
\multicolumn{15}{c}{\textit{Sparse retrieval model}} \\
\hline
BM25 \cite{robertson1995okapi} & - & 19.2 & 27.1 & 14.9 & 12.5 & 13.5 & 16.5 & 15.2 & 24.4 & 7.9 & 6.0 & 13.0 & 6.9 & 14.8 \\
BM25 & \ranker\text{ (3B)} & 34.3 & 44.2 & 15.8 & 18.9 & 15.5 & 20.1 & 21.6 & 23.4 & 10.3 & 6.1 & 10.3 & 6.7 & 18.9 \\
BM25 & \ranker\text{ } w/o expl. (3B) & 15.1 & 20.2 & 12.1 & 10.2 & 11.2 & 13.3 & 12.8 & 19.9 & 6.1 & 4.8 & 10.1 & 5.2 & 11.8 \\
BM25 & \textbf{\ranker\text{ +} instruct (3B)} & 36.0 & 45.0 & 16.3 & 19.8 & 15.3 & 20.3 & 23.7 & 26.9 & 9.0 & 6.6 & 9.1 & 6.2 & 19.5 \\
\hline
\multicolumn{15}{c}{\textit{Dense retrieval models with < 1B parameters}} \\
\hline
TAS-B (66M) \cite{Hofstaetter2021_tasb_dense_retrieval} & - & 2.7 & 10.2 & 6.4 & 5.6 & 7.5 & 8.0 & 4.1 & 24.7 & 14.6 & 8.7 & 7.9 & 1.5 & 8.5 \\
BGE (0.3B) \cite{bge} & - & 12.0 & 24.2 & 16.6 & 17.4 & 12.2 & 9.5 & 13.3 & 26.7 & 5.6 & 6.0 & 13.0 & 6.9 & 13.6 \\
Inst-L (0.3B) \cite{inst} & - & 15.6 & 21.5 & 16.0 & 21.9 & 11.5 & 11.2 & 13.2 & 20.0 & 1.3 & 8.1 & 20.9 & 9.1 & 14.2 \\
GTE-L (0.4B) \cite{li2023towards} & - & 21.0 & 31.1 & 20.5 & 24.3 & 12.6 & 15.9 & 15.3 & 28.3 & 7.3 & 8.3 & 20.3 & 11.6 & 18.0 \\
GTE-L (0.4B) & MiniLM-MARCO (33M) \cite{reimers-gurevych-2019-sentence} & 11.6 & 5.7 & 5.2 & 6.4 & 2.7 & 4.0 & 5.5 & 4.0 & 2.1 & 0.0 & 3.6 & 1.2 & 4.3 \\
GTE-L (0.4B) & ModernBERT-L (0.4B) \cite{warner2024smarterbetterfasterlonger} & 10.8 & 7.1 & 7.2 & 7.4 & 4.1 & 5.8 & 7.0 & 5.8 & 6.8 & 6.4 & 4.6 & 3.8 & 6.4 \\
GTE-L (0.4B) & Llama3.2 (3B) \cite{grattafiori2024llama3herdmodels} & 18.3 & 22.5 & 11.1 & 17.5 & 7.1 & 11.1 & 12.6 & 18.6 & 7.0 & 4.0 & 18.0 & 15.7 & 13.6 \\
GTE-L (0.4B) & Llama3.3 (70B) \cite{grattafiori2024llama3herdmodels} & 29.6 & 37.2 & 23.1 & 30.7 & 13.6 & 22.8 & 20.5 & 23.7 & 18.6 & 7.0 & 23.9 & 23.4 & 22.8 \\
GTE-L (0.4B) & \ranker\text{ (3B)} & \underline{35.2} & \underline{45.7} & \textbf{\underline{24.1}} & 27.4 & \underline{16.1} & 21.8 & \textbf{\underline{20.8}} & 22.0 & 11.7 & \underline{8.7} & 17.4 & 7.5 & 21.5 \\
GTE-L (0.4B) & \ranker\text{ } w/o expl. (3B) & 20.3 & 19.7 & 14.4 & 16.1 & 13.1 & 11.4 & 13.8 & 19.6 & 10.2 & 9.1 & 15.4 & 9.4 & 14.4 \\
GTE-L (0.4B) & \textbf{\ranker\text{ +} instruct (3B)} & 37.0 & 46.5 & 24.8 & 28.8 & 15.8 & 22.1 & 23.0 & 25.3 & 10.2 & 9.5 & 15.4 & 7.0 & 22.1 \\
\hline
\multicolumn{15}{c}{\textit{Dense retrieval models with > 1B parameters}} \\
\hline
E5 (7B) \cite{wang2022text} & - & 18.8 & 26.0 & 15.5 & 15.8 & 16.4 & 9.8 & 18.5 & 28.7 & 4.8 & 7.1 & 26.1 & 26.8 & 17.9 \\
Inst-XL (1.5B) \cite{inst} & - & 21.9 & 34.4 & 22.8 & 27.4 & 17.4 & 19.1 & 18.8 & 27.5 & 5.0 & 8.5 & 15.6 & 5.9 & 18.7 \\
GritLM (7B) \cite{muennighoff2024generative} & - & 25.0 & 32.8 & 19.0 & 19.9 & 17.3 & 11.6 & 18.0 & \textbf{29.8} & 22.0 & 8.8 & 25.1 & 21.1 & 20.9 \\
Qwen1.5 (7B) \cite{bai2023qwentechnicalreport} & - & 30.1 & 38.3 & 17.7 & 23.7 & 13.3 & 22.4 & 14.6 & 25.5 & 8.7 & 14.5 & 27.7 & \textbf{32.8} & 22.4 \\
Qwen1.5 (7B) & MiniLM-MARCO (33M) \cite{reimers-gurevych-2019-sentence} & 9.72 & 6.21 & 6.60 & 6.72 & 3.59 & 5.12 & 6.25 & 5.11 & 6.10 & 5.90 & 4.04 & 3.26 & 5.72 \\
Qwen1.5 (7B) & ModernBERT-L (0.4B) \cite{warner2024smarterbetterfasterlonger} & 11.8 & 8.1 & 8.2 & 8.4 & 5.1 & 6.8 & 8.0 & 6.8 & 7.8 & 7.4 & 5.6 & 4.8 & 7.4 \\
Qwen1.5 (7B) & Llama3.2 (3B) \cite{grattafiori2024llama3herdmodels} & 27.6 & 30.3 & 14.6 & 19.5 & 9.7 & 17.6 & 11.9 & 25.4 & 14.6 & 12.8 & 25.6 & 26.1 & 19.6 \\
Qwen1.5 (7B) & \ranker\text{ (3B)} & 48.5 & 50.6 & 21.7 & 30.3 & \textbf{17.6} & 26.3 & 20.2 & 21.3 & \textbf{26.7} & 12.4 & \textbf{21.7} & 27.4 & 27.1 \\
Qwen1.5 (7B) & \ranker\text{ } w/o expl. (3B) & 21.3 & 25.6 & 15.2 & 16.8 & 13.8 & 16.2 & 15.1 & 22.4 & 11.2 & 10.1 & 16.2 & 10.1 & 16.2 \\
Qwen1.5 (7B) & \textbf{\ranker\text{ +} instruct (3B)} & \textbf{51.2}* & \textbf{51.4}* & 22.4 & \textbf{31.9}* & 17.3 & \textbf{26.6}* & \textbf{22.4}* & 24.5 & 23.1 & \textbf{13.5}* & 19.3 & 25.5 & \textbf{27.4}* \\
\hline
\end{tabular}
\label{tab:retriever_reranker_comparison}
\end{table*}

\begin{table}[h]
\vspace{-0.2cm}
\centering
\footnotesize
\caption{Performance (nDCG@10) of the reranker in various training stages with GTE-large as first-stage retriever.}
\begin{tabular}{l|ccc|cc}
\hline
\textbf{Domain} & \textbf{SFT} & \textbf{RL, t=1} & \textbf{RL, t=2} & \textbf{$\delta$(t=1 vs SFT)} & \textbf{$\delta$(t=2 vs t=1)} \\
\hline
Bio. & \textbf{39.4} & 35.2 & 30.0 & -4.2 & -5.2 \\
Earth.& 42.4 & \textbf{45.7} & 38.4 & \textbf{+3.3} & -7.3 \\
Econ. & 23.2 & \textbf{24.1} & 22.7 & \textbf{+0.9} & -1.4 \\
Psy. & 27.1 & \textbf{27.4} & 25.9 & \textbf{+0.3} & -1.5 \\
Rob. & 14.1 & \textbf{16.1} & 13.6 & \textbf{+2.0} & -2.5 \\
Stack. & \textbf{21.8} & \textbf{21.8} & 18.2 & 0.0 & -3.6 \\
Sus. & 20.6 & \textbf{20.8} & 16.7 & \textbf{+0.2} & -4.1 \\
Leet. & 19.9 & 22.0 & \textbf{25.3} & \textbf{+2.1} & \textbf{+3.3} \\
Pony & 7.4 & 11.7 & \textbf{15.6} & \textbf{+4.3} & \textbf{+3.9} \\
AoPS & 6.9 & \textbf{8.7} & 8.6 & \textbf{+1.8} & -0.1 \\
TheoQ. & 12.8 & 17.4 & \textbf{20.2} & \textbf{+4.6} & \textbf{+2.8} \\
TheoT. & 8.8 & 7.5 & \textbf{9.8} & -1.3 & \textbf{+2.3} \\
\hline
\textbf{Average} & 20.3 & \textbf{21.5} & 20.4 & \textbf{+1.1} & -1.1 \\
\hline
\end{tabular}
\vspace{0.2cm}
\label{tab:ranker_comparison}
\vspace{-0.6cm}
\end{table}

\section{Experiments}
\label{experiments}
\paragraph{Evaluation Data}
Our evaluation uses the BRIGHT benchmark \cite{su2024bright}, which spans diverse domains requiring complex reasoning capabilities. BRIGHT includes seven datasets from StackExchange communities (Biology, Earth Science, Economics, Psychology, Robotics, Stack Overflow, and Sustainable Living), each containing 100-200 expert-validated query-document pairs where relevance is determined by citations in accepted answers. The remaining 5 datasets focus on coding and mathematical reasoning: Pony (syntax documentation pairs), LeetCode (algorithmic problems), TheoremQA (theorem-based questions), AoPS (competition math problems), and Theorem Retrieval (problems paired with ProofWiki statements). In total, BRIGHT contains 1,384 queries with 6.37 positive documents per query on average. The queries are typically long-form questions requiring multi-step reasoning, while positive documents provide critical concepts, theories, or techniques needed to address the queries rather than direct answers.

\paragraph{Experimental Setup} The base LLM for \ranker\text{} is Llama 3.2 3B, while Llama 3.3 70B is our teacher model \cite{grattafiori2024llama3herdmodels}. We use QLoRA \cite{dettmers2024qlora} for parameter-efficient fine-tuning, with a 4-bit quantization of the base model and trainable rank-64 adapters. Due to resource constraints, we limit the context length to 4K tokens. Training is performed on a single A100 GPU with an effective batch size of 16 (batch size 1 with 16 gradient accumulation steps) using the AdamW optimizer with learning rate 2e-4. For the sampling of outputs in the RL stage, we use temperature 1.0 for nucleus sampling, reward threshold $\tau=0.85$, and reward scaling power $m=3$. We perform two epochs of RL training. For the reward model, we use a pretrained Llama 3.1 8B model \cite{liu2024skywork}\footnote{\url{https://hf.co/Skywork/Skywork-Reward-Llama-3.1-8B-v0.2}} that has demonstrated strong performance on RewardBench \cite{rewardbench}. We found that this model has very high agreement with larger open-weight and commercial LLMs in relative comparison of explanation outputs, making it suitable for our training process.

\paragraph{Baselines}
We compare against a diverse set of baseline models: (1) Traditional sparse retrieval using BM25 \cite{robertson1995okapi}; (2) Dense retrievers of varying sizes, from MSMARCO-trained models like TAS-B (66M) \cite{Hofstaetter2021_tasb_dense_retrieval} to recent models like BGE (0.3B) \cite{bge}, Instruction-tuned models Inst-L/XL \cite{inst}, GTE-large (0.4B) \cite{li2023towards}, E5 (7B) \cite{wang2022text}, GritLM (7B) \cite{muennighoff2024generative}, and Qwen1.5 (7B) \cite{bai2023qwentechnicalreport}; (3) Cross-encoder rerankers including MiniLM fine-tuned on MSMARCO \cite{reimers-gurevych-2019-sentence} and ModernBERT-large fine-tuned on our synthetic examples \cite{warner2024smarterbetterfasterlonger}; and (4) Zero-shot LLM rerankers using Llama 3.2 (3B) and Llama 3.3 (70B). These baselines represent the spectrum of current approaches, from lightweight traditional methods to LLMs.



\subsection{Experimental Results} 
Our experimental results, shown in Table \ref{tab:retriever_reranker_comparison}, reveal several key findings about reasoning-intensive ranking:

\textit{\textbf{1. Traditional dense retrievers with small number of parameters or training data fail in reasoning-intensive domains.}} Smaller dense retrievers trained on MSMARCO like TAS-B (66M) perform poorly with only 8.53\% average nDCG@10, highlighting their limitations beyond simple semantic matching. This is particularly evident in reasoning-intensive domains like theorem-based tasks (1.51\% on TheoT) and complex StackExchange queries (2.77\% on Biology). In contrast, larger dense retrievers trained on more diverse data with 100M+ training examples, show significant improvements; GTE-large (400M) achieves 18.0\% and Qwen1.5 (7B) reaches 22.4\% average nDCG@10, demonstrating the importance of model scale and training data for complex retrieval tasks.

\textit{\textbf{2. Explanations are crucial for effective ranking.}} As shown in Table \ref{tab:retriever_reranker_comparison}, our ablation studies reveal that removing the explanation component (rows marked ``w/o expl.'') causes accuracy to drop significantly from 21.5\% to 14.4\% nDCG@10 on average. Traditional BERT re-ranking models that rely purely on semantic matching also show surprisingly poor performance, with ModernBERT-L achieving only 4.57\% average nDCG@10. This shows that the process of generating explanations helps develop better reasoning capabilities compared to approaches that only predict relevance scores directly.

\textbf{\textit{3. Distillation results in small student models with teacher performance.}} Our results also demonstrate that our approach successfully distills complex reasoning capabilities into a compact 3B parameter model, achieving performance comparable to models over 20 times larger (see Llama 3.3 70B in Table~\ref{tab:retriever_reranker_comparison}). When combined with the Qwen1.5 retriever and domain-specific relevance definitions in the ranker's prompt (rows marked with ``+ instruct'' in Table~\ref{tab:retriever_reranker_comparison}), \ranker\text{ }achieves state-of-the-art performance with an average of 27.4\% across all domains reaching the third spot in BRIGHT leaderboard, just below JudgeRank \cite{judgerank2024}, an ensemble of 3 zero-shot LLMs (8B, 70B, and 405B parameters) and a baseline using Llama 70B with query-rewriting with GPT-4. Our 3B parameter model outperforms all other baseline methods on the BRIGHT benchmark, including recent approaches like Reason-to-Rank \cite{r2r2024} (nDCG@5 26.2 vs 19.6) which uses an 8B parameter model.

\textit{\textbf{4. RL improves reasoning for ranking.}} The iterative RL process shows domain-dependent effects, as detailed in Table \ref{tab:ranker_comparison}. While the first iteration leads to broad improvements (+1.1\% nDCG@10 on average), the second iteration reveals an interesting pattern - performance continues to improve in reasoning-intensive domains like mathematics and coding while declining in domains with simpler reasoning requirements. This suggests that additional RL iterations help refine complex reasoning capabilities but may lead to over-fitting in domains where simpler strategies suffice. Table \ref{tab:ranker_comparison} presents detailed results examining the impact of different training stages. The supervised fine-tuning (SFT) stage establishes strong initial performance, particularly in domains like Biology and Earth Science. The first RL iteration shows the largest gains in theoretical domains (TheoQ), coding tasks (Pony, Leetcode), and earth science. The second iteration further improves performance specifically in reasoning-intensive tasks (Leetcode, Pony, TheoQ, TheoT) while showing decline in simpler domains, highlighting the trade-off between specialized reasoning capabilities and general performance.

\section{Conclusions}

This paper presents a novel approach for training compact language models to perform reasoning-intensive document ranking. Our methodology combines knowledge distillation from a large teacher model with reinforcement learning optimization to create efficient yet powerful ranking models that can explain their decisions. Through extensive experimentation we demonstrate that a 3B parameter LLM achieves performance comparable to models over 20 times larger, reaching state-of-the-art results across diverse domains. Dedicating inference-time compute to generate explanations, rather than directly predicting relevance scores, enables more effective reasoning with smaller language models.

Key findings from our work include: (1) the critical role of explanations in developing robust reasoning capabilities for ranking, as shown by significant performance drops when removing the explanation component; (2) the effectiveness of our two-stage training approach, where supervised fine-tuning establishes strong initial performance and targeted reinforcement learning helps refine complex reasoning abilities; and (3) the importance of combining efficient retrievers with reasoning-capable re-rankers, as demonstrated by the strong performance of our GTE-large and Qwen1.5 retriever combinations.

Our results suggest promising directions for future work, including exploring more sophisticated reward modeling approaches, domain adaptation techniques for specialized reasoning tasks, and developing methods to further reduce model size while maintaining reasoning capabilities. Additionally, applying similar weakly-supervised alignment techniques to optimize first-stage retrievers could potentially lead to end-to-end improvements in reasoning-intensive search. The success of our approach in creating efficient, interpretable ranking models opens new possibilities for deploying reasoning-intensive search systems at scale.

\section*{Acknowledgments}
This work was supported in part by the Center for Intelligent Information Retrieval (CIIR), in part by the Office of Naval Research contract number N000142212688, and in part by NSF grant \#2143434. Any opinions, findings and conclusions or recommendations expressed in this material are those of the authors and do not necessarily reflect those of the sponsors.


\bibliographystyle{ACM-Reference-Format}
\balance
\bibliography{references}


\begin{thebibliography}{33}


\ifx \showCODEN    \undefined \def \showCODEN     #1{\unskip}     \fi
\ifx \showDOI      \undefined \def \showDOI       #1{#1}\fi
\ifx \showISBNx    \undefined \def \showISBNx     #1{\unskip}     \fi
\ifx \showISBNxiii \undefined \def \showISBNxiii  #1{\unskip}     \fi
\ifx \showISSN     \undefined \def \showISSN      #1{\unskip}     \fi
\ifx \showLCCN     \undefined \def \showLCCN      #1{\unskip}     \fi
\ifx \shownote     \undefined \def \shownote      #1{#1}          \fi
\ifx \showarticletitle \undefined \def \showarticletitle #1{#1}   \fi
\ifx \showURL      \undefined \def \showURL       {\relax}        \fi
\providecommand\bibfield[2]{#2}
\providecommand\bibinfo[2]{#2}
\providecommand\natexlab[1]{#1}
\providecommand\showeprint[2][]{arXiv:#2}

\bibitem[Bai et~al\mbox{.}(2023)]%
        {bai2023qwentechnicalreport}
\bibfield{author}{\bibinfo{person}{Jinze Bai}, \bibinfo{person}{Shuai Bai}, \bibinfo{person}{Yunfei Chu}, \bibinfo{person}{Zeyu Cui}, \bibinfo{person}{Kai Dang}, \bibinfo{person}{Xiaodong Deng}, \bibinfo{person}{Yang Fan}, \bibinfo{person}{Wenbin Ge}, \bibinfo{person}{Yu Han}, \bibinfo{person}{Fei Huang}, {and} \bibinfo{person}{et al.}} \bibinfo{year}{2023}\natexlab{}.
\newblock \bibinfo{title}{Qwen Technical Report}.
\newblock
\newblock
\showeprint[arxiv]{2309.16609}~[cs.CL]
\urldef\tempurl%
\url{https://arxiv.org/abs/2309.16609}
\showURL{%
\tempurl}


\bibitem[Burges(2010)]%
        {burges2010ranknet}
\bibfield{author}{\bibinfo{person}{Christopher~JC Burges}.} \bibinfo{year}{2010}\natexlab{}.
\newblock \showarticletitle{From ranknet to lambdarank to lambdamart: An overview}.
\newblock \bibinfo{journal}{\emph{Learning}} \bibinfo{volume}{11}, \bibinfo{number}{23-581} (\bibinfo{year}{2010}), \bibinfo{pages}{81}.
\newblock


\bibitem[Craswell et~al\mbox{.}(2020)]%
        {craswell2020overview}
\bibfield{author}{\bibinfo{person}{Nick Craswell}, \bibinfo{person}{Bhaskar Mitra}, \bibinfo{person}{Emine Yilmaz}, \bibinfo{person}{Daniel Campos}, {and} \bibinfo{person}{Ellen~M Voorhees}.} \bibinfo{year}{2020}\natexlab{}.
\newblock \showarticletitle{Overview of the TREC 2019 deep learning track}.
\newblock \bibinfo{journal}{\emph{arXiv preprint arXiv:2003.07820}} (\bibinfo{year}{2020}).
\newblock


\bibitem[Craswell et~al\mbox{.}(2021)]%
        {craswell2021trec}
\bibfield{author}{\bibinfo{person}{Nick Craswell}, \bibinfo{person}{Bhaskar Mitra}, \bibinfo{person}{Emine Yilmaz}, \bibinfo{person}{Daniel Campos}, \bibinfo{person}{Ellen~M Voorhees}, {and} \bibinfo{person}{Ian Soboroff}.} \bibinfo{year}{2021}\natexlab{}.
\newblock \showarticletitle{TREC deep learning track: Reusable test collections in the large data regime}. In \bibinfo{booktitle}{\emph{Proceedings of the 44th international ACM SIGIR conference on research and development in information retrieval}}. \bibinfo{pages}{2369--2375}.
\newblock


\bibitem[DeepSeek-AI et~al\mbox{.}(2025)]%
        {deepseekr1}
\bibfield{author}{\bibinfo{person}{DeepSeek-AI}, \bibinfo{person}{Daya Guo}, \bibinfo{person}{Dejian Yang}, \bibinfo{person}{Haowei Zhang}, \bibinfo{person}{Junxiao Song}, \bibinfo{person}{Ruoyu Zhang}, \bibinfo{person}{Runxin Xu}, \bibinfo{person}{Qihao Zhu}, \bibinfo{person}{Shirong Ma}, \bibinfo{person}{Peiyi Wang}, \bibinfo{person}{Xiao Bi}, \bibinfo{person}{Xiaokang Zhang}, \bibinfo{person}{Xingkai Yu}, \bibinfo{person}{Yu Wu}, \bibinfo{person}{Z.~F. Wu}, \bibinfo{person}{Zhibin Gou}, \bibinfo{person}{Zhihong Shao}, \bibinfo{person}{Zhuoshu Li}, \bibinfo{person}{Ziyi Gao}, \bibinfo{person}{Aixin Liu}, \bibinfo{person}{Bing Xue}, {and} \bibinfo{person}{Bingxuan~Wang et al.}} \bibinfo{year}{2025}\natexlab{}.
\newblock \bibinfo{title}{DeepSeek-R1: Incentivizing Reasoning Capability in LLMs via Reinforcement Learning}.
\newblock
\newblock
\showeprint[arxiv]{2501.12948}~[cs.CL]
\urldef\tempurl%
\url{https://arxiv.org/abs/2501.12948}
\showURL{%
\tempurl}


\bibitem[Dettmers et~al\mbox{.}(2024)]%
        {dettmers2024qlora}
\bibfield{author}{\bibinfo{person}{Tim Dettmers}, \bibinfo{person}{Artidoro Pagnoni}, \bibinfo{person}{Ari Holtzman}, {and} \bibinfo{person}{Luke Zettlemoyer}.} \bibinfo{year}{2024}\natexlab{}.
\newblock \showarticletitle{Qlora: Efficient finetuning of quantized llms}.
\newblock \bibinfo{journal}{\emph{Advances in Neural Information Processing Systems}}  \bibinfo{volume}{36} (\bibinfo{year}{2024}).
\newblock


\bibitem[Devlin et~al\mbox{.}(2018)]%
        {devlin2018bert}
\bibfield{author}{\bibinfo{person}{Jacob Devlin}, \bibinfo{person}{Ming-Wei Chang}, \bibinfo{person}{Kenton Lee}, {and} \bibinfo{person}{Kristina Toutanova}.} \bibinfo{year}{2018}\natexlab{}.
\newblock \showarticletitle{BERT: Pre-training of Deep Bidirectional Transformers for Language Understanding}.
\newblock \bibinfo{journal}{\emph{arXiv preprint arXiv:1810.04805}} (\bibinfo{year}{2018}).
\newblock


\bibitem[Gao et~al\mbox{.}(2021)]%
        {gao2021rethink}
\bibfield{author}{\bibinfo{person}{Luyu Gao}, \bibinfo{person}{Zhuyun Dai}, {and} \bibinfo{person}{Jamie Callan}.} \bibinfo{year}{2021}\natexlab{}.
\newblock \showarticletitle{Rethink training of BERT rerankers in multi-stage retrieval pipeline}. In \bibinfo{booktitle}{\emph{Advances in Information Retrieval: 43rd European Conference on IR Research, ECIR 2021}}. Springer, \bibinfo{pages}{280--286}.
\newblock


\bibitem[Grattafiori et~al\mbox{.}(2024)]%
        {grattafiori2024llama3herdmodels}
\bibfield{author}{\bibinfo{person}{Aaron Grattafiori}, \bibinfo{person}{Abhimanyu Dubey}, \bibinfo{person}{Abhinav Jauhri}, \bibinfo{person}{Abhinav Pandey}, \bibinfo{person}{Abhishek Kadian}, {and} \bibinfo{person}{Ahmad Al-Dahle et al.}} \bibinfo{year}{2024}\natexlab{}.
\newblock \bibinfo{title}{The Llama 3 Herd of Models}.
\newblock
\newblock
\showeprint[arxiv]{2407.21783}~[cs.AI]
\urldef\tempurl%
\url{https://arxiv.org/abs/2407.21783}
\showURL{%
\tempurl}


\bibitem[Hinton et~al\mbox{.}(2015)]%
        {hinton2015distillingknowledgeneuralnetwork}
\bibfield{author}{\bibinfo{person}{Geoffrey Hinton}, \bibinfo{person}{Oriol Vinyals}, {and} \bibinfo{person}{Jeff Dean}.} \bibinfo{year}{2015}\natexlab{}.
\newblock \bibinfo{title}{Distilling the Knowledge in a Neural Network}.
\newblock
\newblock
\showeprint[arxiv]{1503.02531}~[stat.ML]
\urldef\tempurl%
\url{https://arxiv.org/abs/1503.02531}
\showURL{%
\tempurl}


\bibitem[Hofst{\"a}tter et~al\mbox{.}(2021)]%
        {Hofstaetter2021_tasb_dense_retrieval}
\bibfield{author}{\bibinfo{person}{Sebastian Hofst{\"a}tter}, \bibinfo{person}{Sheng-Chieh Lin}, \bibinfo{person}{Jheng-Hong Yang}, \bibinfo{person}{Jimmy Lin}, {and} \bibinfo{person}{Allan Hanbury}.} \bibinfo{year}{2021}\natexlab{}.
\newblock \showarticletitle{{Efficiently Teaching an Effective Dense Retriever with Balanced Topic Aware Sampling}}. In \bibinfo{booktitle}{\emph{Proc. of SIGIR}}.
\newblock


\bibitem[Ji et~al\mbox{.}(2024)]%
        {r2r2024}
\bibfield{author}{\bibinfo{person}{Yuelyu Ji}, \bibinfo{person}{Zhuochun Li}, \bibinfo{person}{Rui Meng}, {and} \bibinfo{person}{Daqing He}.} \bibinfo{year}{2024}\natexlab{}.
\newblock \showarticletitle{ReasoningRank: Teaching Student Models to Rank through Reasoning-Based Knowledge Distillation}.
\newblock  (\bibinfo{year}{2024}).
\newblock
\showeprint[arxiv]{2410.05168}~[cs.CL]
\urldef\tempurl%
\url{https://arxiv.org/abs/2410.05168}
\showURL{%
\tempurl}


\bibitem[Killingback et~al\mbox{.}(2025)]%
        {hypencoder}
\bibfield{author}{\bibinfo{person}{Julian Killingback}, \bibinfo{person}{Hansi Zeng}, {and} \bibinfo{person}{Hamed Zamani}.} \bibinfo{year}{2025}\natexlab{}.
\newblock \bibinfo{title}{Hypencoder: Hypernetworks for Information Retrieval}.
\newblock
\newblock
\showeprint[arxiv]{2502.05364}~[cs.IR]
\urldef\tempurl%
\url{https://arxiv.org/abs/2502.05364}
\showURL{%
\tempurl}


\bibitem[Lambert et~al\mbox{.}(2024)]%
        {rewardbench}
\bibfield{author}{\bibinfo{person}{Nathan Lambert}, \bibinfo{person}{Valentina Pyatkin}, \bibinfo{person}{Jacob Morrison}, \bibinfo{person}{LJ Miranda}, \bibinfo{person}{Bill~Yuchen Lin}, \bibinfo{person}{Khyathi Chandu}, \bibinfo{person}{Nouha Dziri}, \bibinfo{person}{Sachin Kumar}, \bibinfo{person}{Tom Zick}, \bibinfo{person}{Yejin Choi}, \bibinfo{person}{Noah~A. Smith}, {and} \bibinfo{person}{Hannaneh Hajishirzi}.} \bibinfo{year}{2024}\natexlab{}.
\newblock \bibinfo{title}{RewardBench: Evaluating Reward Models for Language Modeling}.
\newblock
\newblock
\showeprint[arxiv]{2403.13787}~[cs.LG]
\urldef\tempurl%
\url{https://arxiv.org/abs/2403.13787}
\showURL{%
\tempurl}


\bibitem[Li et~al\mbox{.}(2023)]%
        {li2023towards}
\bibfield{author}{\bibinfo{person}{Zehan Li}, \bibinfo{person}{Xin Zhang}, \bibinfo{person}{Yanzhao Zhang}, \bibinfo{person}{Dingkun Long}, \bibinfo{person}{Pengjun Xie}, {and} \bibinfo{person}{Meishan Zhang}.} \bibinfo{year}{2023}\natexlab{}.
\newblock \showarticletitle{Towards general text embeddings with multi-stage contrastive learning}.
\newblock \bibinfo{journal}{\emph{arXiv preprint arXiv:2308.03281}} (\bibinfo{year}{2023}).
\newblock


\bibitem[Liu et~al\mbox{.}(2024)]%
        {liu2024skywork}
\bibfield{author}{\bibinfo{person}{Chris~Yuhao Liu}, \bibinfo{person}{Liang Zeng}, \bibinfo{person}{Jiacai Liu}, \bibinfo{person}{Rui Yan}, \bibinfo{person}{Jujie He}, \bibinfo{person}{Chaojie Wang}, \bibinfo{person}{Shuicheng Yan}, \bibinfo{person}{Yang Liu}, {and} \bibinfo{person}{Yahui Zhou}.} \bibinfo{year}{2024}\natexlab{}.
\newblock \showarticletitle{Skywork-Reward: Bag of Tricks for Reward Modeling in LLMs}.
\newblock \bibinfo{journal}{\emph{arXiv preprint arXiv:2410.18451}} (\bibinfo{year}{2024}).
\newblock


\bibitem[Liu(2010)]%
        {ltr}
\bibfield{author}{\bibinfo{person}{Tie-Yan Liu}.} \bibinfo{year}{2010}\natexlab{}.
\newblock \showarticletitle{Learning to rank for information retrieval}. In \bibinfo{booktitle}{\emph{Proceedings of the 33rd International ACM SIGIR Conference on Research and Development in Information Retrieval}} (Geneva, Switzerland) \emph{(\bibinfo{series}{SIGIR '10})}. \bibinfo{publisher}{Association for Computing Machinery}, \bibinfo{address}{New York, NY, USA}, \bibinfo{pages}{904}.
\newblock
\showISBNx{9781450301534}
\urldef\tempurl%
\url{https://doi.org/10.1145/1835449.1835676}
\showDOI{\tempurl}


\bibitem[Muennighoff et~al\mbox{.}(2024)]%
        {muennighoff2024generative}
\bibfield{author}{\bibinfo{person}{Niklas Muennighoff}, \bibinfo{person}{Hongjin Su}, \bibinfo{person}{Liang Wang}, \bibinfo{person}{Nan Yang}, \bibinfo{person}{Furu Wei}, \bibinfo{person}{Tao Yu}, \bibinfo{person}{Amanpreet Singh}, {and} \bibinfo{person}{Douwe Kiela}.} \bibinfo{year}{2024}\natexlab{}.
\newblock \showarticletitle{Generative representational instruction tuning}.
\newblock \bibinfo{journal}{\emph{arXiv preprint arXiv:2402.09906}} (\bibinfo{year}{2024}).
\newblock


\bibitem[Nguyen et~al\mbox{.}(2016)]%
        {DBLP:journals/corr/NguyenRSGTMD16}
\bibfield{author}{\bibinfo{person}{Tri Nguyen}, \bibinfo{person}{Mir Rosenberg}, \bibinfo{person}{Xia Song}, \bibinfo{person}{Jianfeng Gao}, \bibinfo{person}{Saurabh Tiwary}, \bibinfo{person}{Rangan Majumder}, {and} \bibinfo{person}{Li Deng}.} \bibinfo{year}{2016}\natexlab{}.
\newblock \showarticletitle{{MS} {MARCO:} {A} Human Generated MAchine Reading COmprehension Dataset}.
\newblock \bibinfo{journal}{\emph{CoRR}}  \bibinfo{volume}{abs/1611.09268} (\bibinfo{year}{2016}).
\newblock
\showeprint[arxiv]{1611.09268}
\urldef\tempurl%
\url{http://arxiv.org/abs/1611.09268}
\showURL{%
\tempurl}


\bibitem[Niu et~al\mbox{.}(2024)]%
        {judgerank2024}
\bibfield{author}{\bibinfo{person}{Tong Niu}, \bibinfo{person}{Shafiq Joty}, \bibinfo{person}{Ye Liu}, \bibinfo{person}{Caiming Xiong}, \bibinfo{person}{Yingbo Zhou}, {and} \bibinfo{person}{Semih Yavuz}.} \bibinfo{year}{2024}\natexlab{}.
\newblock \showarticletitle{JudgeRank: Leveraging Large Language Models for Reasoning-Intensive Reranking}.
\newblock  (\bibinfo{year}{2024}).
\newblock
\showeprint[arxiv]{2411.00142}~[cs.CL]
\urldef\tempurl%
\url{https://arxiv.org/abs/2411.00142}
\showURL{%
\tempurl}


\bibitem[Nogueira and Cho(2019)]%
        {nogueira2019passage}
\bibfield{author}{\bibinfo{person}{Rodrigo Nogueira} {and} \bibinfo{person}{Kyunghyun Cho}.} \bibinfo{year}{2019}\natexlab{}.
\newblock \showarticletitle{Passage Re-ranking with BERT}.
\newblock \bibinfo{journal}{\emph{arXiv preprint arXiv:1901.04085}} (\bibinfo{year}{2019}).
\newblock


\bibitem[Nogueira et~al\mbox{.}(2019)]%
        {nogueira2019multi}
\bibfield{author}{\bibinfo{person}{Rodrigo Nogueira}, \bibinfo{person}{Wei Yang}, \bibinfo{person}{Kyunghyun Cho}, {and} \bibinfo{person}{Jimmy Lin}.} \bibinfo{year}{2019}\natexlab{}.
\newblock \showarticletitle{Multi-stage document ranking with BERT}.
\newblock \bibinfo{journal}{\emph{arXiv preprint arXiv:1910.14424}} (\bibinfo{year}{2019}).
\newblock


\bibitem[Pradeep et~al\mbox{.}(2023a)]%
        {pradeep2023rankvicuna}
\bibfield{author}{\bibinfo{person}{Ronak Pradeep}, \bibinfo{person}{Sahel Sharifymoghaddam}, {and} \bibinfo{person}{Jimmy Lin}.} \bibinfo{year}{2023}\natexlab{a}.
\newblock \showarticletitle{RankVicuna: Zero-shot listwise document reranking with open-source large language models}.
\newblock \bibinfo{journal}{\emph{arXiv preprint arXiv:2309.15088}} (\bibinfo{year}{2023}).
\newblock


\bibitem[Pradeep et~al\mbox{.}(2023b)]%
        {pradeep2023rankzephyr}
\bibfield{author}{\bibinfo{person}{Ronak Pradeep}, \bibinfo{person}{Sahel Sharifymoghaddam}, {and} \bibinfo{person}{Jimmy Lin}.} \bibinfo{year}{2023}\natexlab{b}.
\newblock \showarticletitle{RankZephyr: Effective and Robust Zero-Shot Listwise Reranking is a Breeze!}
\newblock \bibinfo{journal}{\emph{arXiv preprint arXiv:2312.02724}} (\bibinfo{year}{2023}).
\newblock


\bibitem[Reimers and Gurevych(2019)]%
        {reimers-gurevych-2019-sentence}
\bibfield{author}{\bibinfo{person}{Nils Reimers} {and} \bibinfo{person}{Iryna Gurevych}.} \bibinfo{year}{2019}\natexlab{}.
\newblock \showarticletitle{Sentence-{BERT}: Sentence Embeddings using {S}iamese {BERT}-Networks}. In \bibinfo{booktitle}{\emph{Proceedings of the 2019 Conference on Empirical Methods in Natural Language Processing and the 9th International Joint Conference on Natural Language Processing (EMNLP-IJCNLP)}}, \bibfield{editor}{\bibinfo{person}{Kentaro Inui}, \bibinfo{person}{Jing Jiang}, \bibinfo{person}{Vincent Ng}, {and} \bibinfo{person}{Xiaojun Wan}} (Eds.). \bibinfo{publisher}{Association for Computational Linguistics}, \bibinfo{address}{Hong Kong, China}, \bibinfo{pages}{3982--3992}.
\newblock
\urldef\tempurl%
\url{https://doi.org/10.18653/v1/D19-1410}
\showDOI{\tempurl}


\bibitem[Robertson et~al\mbox{.}(1995)]%
        {robertson1995okapi}
\bibfield{author}{\bibinfo{person}{Stephen Robertson}, \bibinfo{person}{S. Walker}, \bibinfo{person}{S. Jones}, \bibinfo{person}{M.~M. Hancock-Beaulieu}, {and} \bibinfo{person}{M. Gatford}.} \bibinfo{year}{1995}\natexlab{}.
\newblock \showarticletitle{Okapi at TREC-3}. In \bibinfo{booktitle}{\emph{Overview of the Third Text REtrieval Conference (TREC-3)} (\bibinfo{edition}{overview of the third text retrieval conference (trec–3)} ed.)}. \bibinfo{publisher}{Gaithersburg, MD: NIST}, \bibinfo{pages}{109--126}.
\newblock
\urldef\tempurl%
\url{https://www.microsoft.com/en-us/research/publication/okapi-at-trec-3/}
\showURL{%
\tempurl}


\bibitem[Samarinas et~al\mbox{.}(2021)]%
        {samarinas-etal-2021-improving}
\bibfield{author}{\bibinfo{person}{Chris Samarinas}, \bibinfo{person}{Wynne Hsu}, {and} \bibinfo{person}{Mong~Li Lee}.} \bibinfo{year}{2021}\natexlab{}.
\newblock \showarticletitle{Improving Evidence Retrieval for Automated Explainable Fact-Checking}. In \bibinfo{booktitle}{\emph{Proceedings of the 2021 Conference of the North American Chapter of the Association for Computational Linguistics: Human Language Technologies: Demonstrations}}, \bibfield{editor}{\bibinfo{person}{Avi Sil} {and} \bibinfo{person}{Xi~Victoria Lin}} (Eds.). \bibinfo{publisher}{Association for Computational Linguistics}, \bibinfo{address}{Online}, \bibinfo{pages}{84--91}.
\newblock
\urldef\tempurl%
\url{https://doi.org/10.18653/v1/2021.naacl-demos.10}
\showDOI{\tempurl}


\bibitem[Su et~al\mbox{.}(2023)]%
        {inst}
\bibfield{author}{\bibinfo{person}{Hongjin Su}, \bibinfo{person}{Weijia Shi}, \bibinfo{person}{Jungo Kasai}, \bibinfo{person}{Yizhong Wang}, \bibinfo{person}{Yushi Hu}, \bibinfo{person}{Mari Ostendorf}, \bibinfo{person}{Wen-tau Yih}, \bibinfo{person}{Noah~A. Smith}, \bibinfo{person}{Luke Zettlemoyer}, {and} \bibinfo{person}{Tao Yu}.} \bibinfo{year}{2023}\natexlab{}.
\newblock \showarticletitle{One Embedder, Any Task: Instruction-Finetuned Text Embeddings}. In \bibinfo{booktitle}{\emph{Findings of the Association for Computational Linguistics: ACL 2023}}, \bibfield{editor}{\bibinfo{person}{Anna Rogers}, \bibinfo{person}{Jordan Boyd-Graber}, {and} \bibinfo{person}{Naoaki Okazaki}} (Eds.). \bibinfo{publisher}{Association for Computational Linguistics}, \bibinfo{address}{Toronto, Canada}, \bibinfo{pages}{1102--1121}.
\newblock
\urldef\tempurl%
\url{https://doi.org/10.18653/v1/2023.findings-acl.71}
\showDOI{\tempurl}


\bibitem[SU et~al\mbox{.}(2025)]%
        {su2024bright}
\bibfield{author}{\bibinfo{person}{Hongjin SU}, \bibinfo{person}{Howard Yen}, \bibinfo{person}{Mengzhou Xia}, \bibinfo{person}{Weijia Shi}, \bibinfo{person}{Niklas Muennighoff}, \bibinfo{person}{Han yu Wang}, \bibinfo{person}{Liu Haisu}, \bibinfo{person}{Quan Shi}, \bibinfo{person}{Zachary~S Siegel}, \bibinfo{person}{Michael Tang}, \bibinfo{person}{Ruoxi Sun}, \bibinfo{person}{Jinsung Yoon}, \bibinfo{person}{Sercan~O Arik}, \bibinfo{person}{Danqi Chen}, {and} \bibinfo{person}{Tao Yu}.} \bibinfo{year}{2025}\natexlab{}.
\newblock \showarticletitle{{BRIGHT}: A Realistic and Challenging Benchmark for Reasoning-Intensive Retrieval}. In \bibinfo{booktitle}{\emph{The Thirteenth International Conference on Learning Representations}}.
\newblock
\urldef\tempurl%
\url{https://openreview.net/forum?id=ykuc5q381b}
\showURL{%
\tempurl}


\bibitem[Sun et~al\mbox{.}(2023)]%
        {sun2023chatgpt}
\bibfield{author}{\bibinfo{person}{Weiwei Sun}, \bibinfo{person}{Lingyong Yan}, \bibinfo{person}{Xinyu Ma}, \bibinfo{person}{Shuaiqiang Wang}, \bibinfo{person}{Pengjie Ren}, \bibinfo{person}{Zhumin Chen}, \bibinfo{person}{Dawei Yin}, {and} \bibinfo{person}{Zhaochun Ren}.} \bibinfo{year}{2023}\natexlab{}.
\newblock \showarticletitle{Is ChatGPT good at search? Investigating large language models as re-ranking agents}.
\newblock \bibinfo{journal}{\emph{arXiv preprint arXiv:2304.09542}} (\bibinfo{year}{2023}).
\newblock


\bibitem[Wang et~al\mbox{.}(2022)]%
        {wang2022text}
\bibfield{author}{\bibinfo{person}{Liang Wang}, \bibinfo{person}{Nan Yang}, \bibinfo{person}{Xiaolong Huang}, \bibinfo{person}{Binxing Jiao}, \bibinfo{person}{Linjun Yang}, \bibinfo{person}{Daxin Jiang}, \bibinfo{person}{Rangan Majumder}, {and} \bibinfo{person}{Furu Wei}.} \bibinfo{year}{2022}\natexlab{}.
\newblock \showarticletitle{Text embeddings by weakly-supervised contrastive pre-training}.
\newblock \bibinfo{journal}{\emph{arXiv preprint arXiv:2212.03533}} (\bibinfo{year}{2022}).
\newblock


\bibitem[Warner et~al\mbox{.}(2024)]%
        {warner2024smarterbetterfasterlonger}
\bibfield{author}{\bibinfo{person}{Benjamin Warner}, \bibinfo{person}{Antoine Chaffin}, \bibinfo{person}{Benjamin Clavié}, \bibinfo{person}{Orion Weller}, \bibinfo{person}{Oskar Hallström}, \bibinfo{person}{Said Taghadouini}, \bibinfo{person}{Alexis Gallagher}, \bibinfo{person}{Raja Biswas}, \bibinfo{person}{Faisal Ladhak}, \bibinfo{person}{Tom Aarsen}, \bibinfo{person}{Nathan Cooper}, \bibinfo{person}{Griffin Adams}, \bibinfo{person}{Jeremy Howard}, {and} \bibinfo{person}{Iacopo Poli}.} \bibinfo{year}{2024}\natexlab{}.
\newblock \bibinfo{title}{Smarter, Better, Faster, Longer: A Modern Bidirectional Encoder for Fast, Memory Efficient, and Long Context Finetuning and Inference}.
\newblock
\newblock
\showeprint[arxiv]{2412.13663}~[cs.CL]
\urldef\tempurl%
\url{https://arxiv.org/abs/2412.13663}
\showURL{%
\tempurl}


\bibitem[Xiao et~al\mbox{.}(2024)]%
        {bge}
\bibfield{author}{\bibinfo{person}{Shitao Xiao}, \bibinfo{person}{Zheng Liu}, \bibinfo{person}{Peitian Zhang}, \bibinfo{person}{Niklas Muennighoff}, \bibinfo{person}{Defu Lian}, {and} \bibinfo{person}{Jian-Yun Nie}.} \bibinfo{year}{2024}\natexlab{}.
\newblock \showarticletitle{C-Pack: Packed Resources For General Chinese Embeddings}. In \bibinfo{booktitle}{\emph{Proceedings of the 47th International ACM SIGIR Conference on Research and Development in Information Retrieval}} (Washington DC, USA) \emph{(\bibinfo{series}{SIGIR '24})}. \bibinfo{publisher}{Association for Computing Machinery}, \bibinfo{address}{New York, NY, USA}, \bibinfo{pages}{641–649}.
\newblock
\showISBNx{9798400704314}
\urldef\tempurl%
\url{https://doi.org/10.1145/3626772.3657878}
\showDOI{\tempurl}


\end{thebibliography}


\end{document}